\documentclass[a4paper,12pt]{article}
\usepackage[dvips]{color}
\usepackage{multicol}
\usepackage{amsmath}
\usepackage{graphicx}
\usepackage[latin1]{inputenc}
\date{}
\begin{document}

\title{Explicit expression for Lindhard dielectric function at finite temperature\\}

\author{A. V. Andrade-Neto\\
{\small \emph{Departamento de F\'{\i}sica, Universidade Estadual de Feira de
Santana }}\\
{\small \emph{ Feira de Santana, Bahia, Brasil}}\\
{\small \emph{E-mail: aneto@uefs.br}}}
\maketitle
\begin{abstract}

In this work, within the scope of the Lindhard dielectric function formalism for the homogeneous electron gas, explicit expressions for the real and imaginary parts are calculated for finite temperature. An application to Raman scattering in n-type GaAs is presented  to highlight the power of the method.

\textbf{keywords}: dielectric function, Lindhard function, Raman scattering.
\end{abstract}

\section{Introduction}
\indent
In dealing electrons in metal and semiconductors it is useful to utilize a model in which we neglect the discrete lattice structure of the ions in the solid, which are smeared out to a uniform background. This positive background sustain the electrical neutrality of the system. On the other hand electrons muttually interact by Coulomb's force.

We may calculate a wide variety of properties of the interacting electron gas from the knowledge of the frequency and wavevector dependent dielectric function $\epsilon(\vec{Q},\omega)$, e.g., collective excitations like plasmon and screening effects. Since the pioneering work of Lindhard \cite{Lindhard} a large number of papers has gone alternative expressions for $\epsilon(\vec{Q},\omega)$.

In general the Lindhard dielectric function has been calculated for the degenerate case ($T=0$) \cite{Mahan} and for the case at high temperature \cite{Luzzi,Neto}. For the cases which cover a wide range of temperatures, neither the degenerate limit nor the classical Maxwellian limit are good approximations, and therefore analytical expressions for dielectric function to arbitrary temperature is necessary. In this work the real and imaginary parts for the Lindhard dielectric function are  analytically calculated. The expression for the imaginary part is valid for any finite temperature. The expression for the real part is good for all except for low-temperature limit $T \rightarrow 0$.

\section{Analytical expressions for the Lindhard dielectric function}

The Lindhard dielectric function can be written in the form

\begin{equation}\label{fdielcompacto}
\epsilon(\vec{Q},\omega)=\epsilon_{\infty}-\frac{4 \pi e^2}{Q^2} \chi(\vec{Q},\omega) ~.
\end{equation}
where $\epsilon_{\infty}$  is the background dielectric constant contribution of the inner electrons in the ion core, $e$ is the elementary charge and
$\chi(\vec{Q},\omega)$ is the Lindhard function

\begin{equation}\label{fdielcompacto}
\chi(\vec{Q},\omega)=\frac{1}{4 \pi^3}
\int d^3k
\frac{f(\vec{k}+\vec{Q})-f(\vec{k})}{E(\vec{k}+\vec{Q})-E(\vec{k})
-\hbar(\omega+\imath s)}~.
\end{equation}
where  $f(\vec{k})$ is the Fermi-Dirac distribution function, $E(k)=\hbar^2 k^2/2m^*$ is the electron energy, $s$ is a positive infinitesimal which is taken in the limit of going to $+0$ to produce the real, $\epsilon_{1}(Q,\omega)$, and imaginary parts, $\epsilon_{2}(Q,\omega)$, of $\epsilon(\vec{Q},\omega)$.
Using the Dirac identity

\begin{equation}\label{eq17}
\lim_{s\rightarrow 0} \frac{1}{X +\imath s}=p.v. \frac{1}{X}-\imath \pi \delta(X)~,
\end{equation}
where $p.v.$ denotes the principal value of an integral and $\delta(X)$ is the Dirac delta,  we can obtain the real and imaginary parts of the dielectric function. Using spherical coordinates,$(k,\theta, \phi$), after integration in the angular part we write the real part as

\begin{equation}\label{FQa}
\epsilon_{1}(Q,\omega)=\epsilon_{\infty} - \frac{2 e^2 m^*}{\hbar^2 Q^3}\int_{0}^{\infty}k f(k)\left[ \ln\left| \frac{k-q_{1}}{k+q_{1}}\right|+
\ln\left|\frac{k-q_{2}}{k+q_{2}}\right| \right] dk
\end{equation}
where $q_{1}$ and $q_{2}$ are given by

\begin{equation}
q_{1}=\frac{Q}{2}-\frac{m^* \omega}{\hbar Q}
\end{equation}

\begin{equation}
q_{2}=\frac{Q}{2}+\frac{m^* \omega}{\hbar Q}
\end{equation}

and

\begin{equation}\label{epsoimag1}
\epsilon_{2}(Q,\omega)=\frac{2 e^2 m^*}{\hbar^2 Q^3} \int_{q_{1}}^{q_{2}} k f(k) dk ~,
\end{equation}
for the imaginary part.

So far we have not yet made use of the Fermi-Dirac function. At finite temperature, $T$, the occupation numbers are given by the Fermi-Dirac distribution

\begin{equation}\label{Fermi}
f(E(k))=\frac{1}{\exp{[\beta(E(k)-\mu)]}+1}~,
\end{equation}
where $\beta$ is $1/k_{B}T$  and $\mu$ is the chemical potential.

The integral in Eq.(\ref{FQa}) cannot be evaluated analytically. However, assuming that $e^{-\beta \mu}>> 1$, the Fermi function, Eq. (\ref{Fermi}), can be approximately by $f(k)\approx e^{\beta \mu} e^{-\beta E(k)}$. In this case, Eq.(\ref{FQa}) yields

\begin{equation}
\epsilon_{1}(Q,\omega)=\epsilon_{\infty}-\frac{2 e^2 m^*}{\hbar^2 Q^3} e^{\beta \mu} \int_{0}^{\infty}k e^{-\beta E(k)}\left[ \ln\left| \frac{k-q_{1}}{k+q_{1}}\right|+
\ln\left|\frac{k-q_{2}}{k+q_{2}}\right| \right] dk
\end{equation}

In the above equation we have two integrals of kind

\begin{equation}\label{Int}
I=\int_{0}^{\infty}x \exp[-\xi^2 x^2] \ln\left| \frac{x-q}{x+q}\right| dx
\end{equation}
where $\xi^2=\beta \hbar^2/(2m^*)$. The solution of Eq.(\ref{Int}) is given by

\begin{equation}
I= -\frac{\sqrt{\pi}}{\xi^2}\exp(-y^2)\int_{0}^{y} \exp(x^2)dx=-\frac{\sqrt{\pi}}{\xi^2}D(y)
\end{equation}
where $D(y)=\exp(-y^2)\int_{0}^{y} \exp(x^2)dx$ is the Dawson's integral \cite{Abramowitz}.

After some algebraic manipulations we arrive at the following expression for the real  part

\begin{equation}\label{DFreal}
\epsilon_{1}(Q,\omega, T)=\epsilon_{\infty}+\frac{4 e^2 m^2}{\pi^{1/2}Q^3 \hbar^4 \beta}e^{\beta \mu}[D(y_{1})+D(y_{2})]~.
\end{equation}

Here

\begin{equation}
y_{1}^{2}=\frac{\beta \hbar^2}{2 m^*}\left(\frac{Q}{2}+\frac{m^* \omega}{\hbar Q}\right)^2~,
\end{equation}
and

\begin{equation}
y_{2}^{2}=\frac{\beta \hbar^2}{2 m^*}\left(\frac{Q}{2}-\frac{m^* \omega}{\hbar Q}\right)^2~.
\end{equation}

For imaginary part, integrating the  Eq. (\ref{epsoimag1}) we obtain that

\begin{equation}\label{Imaggeral}
\epsilon_{2}(Q,\omega, T)=\frac{2 e^2 (m^*)^2}{\hbar^3 Q^3} \left[\omega -\frac{1}{\hbar \beta}
\ln{\left(\frac{1+e^{-\beta \mu}e^{y_{1}^2}}{1+e^{-\beta \mu}e^{y_{2}^2}}\right)}\right] ~.
\end{equation}

Rearranging Eq.(\ref{Imaggeral}), we have

\begin{equation}\label{Imaggeralbb}
\epsilon_{2}(Q,\omega, T)=\frac{2 e^2 (m^*)^2}{\hbar^4 Q^3 \beta}
\ln{\left(\frac{1+e^{\beta \mu}e^{-y_{2}^2}}{1+e^{\beta \mu}e^{-y_{1}^2}}\right)} ~.
\end{equation}

The Eq.(\ref{Imaggeralbb}) is valid for any finite temperature.

These expressions for $\epsilon_{1}(Q,\omega, T)$ and $\epsilon_{2}(Q,\omega, T)$ constitute the principal result of this work.

\subsection{Important limiting cases}

To appreciate our results, we discus some important limiting cases. We will consider first the upper temperature limit ($\beta \rightarrow 0$). In this case we have

\begin{equation}
e^{\beta \mu}=2^{1/2}\hbar^3 n \left(\frac{ \pi \beta}{m}\right)^{3/2}~,
\end{equation}
where $n$ is the concentration. Inserting this into Eq. (\ref{DFreal}) we get

\begin{equation}\label{DFreala}
\epsilon_{1}(Q,\omega, T)=\epsilon_{\infty}\left[1+\left(\frac{2 m}{\beta}\right)^{1/2}\frac{k_{DH}^2}{\hbar Q^3}[D(y_{1})+D(y_{2})]\right]~.
\end{equation}
where

\begin{equation}\label{Debye}
k_{DH}^2=\frac{4 \pi e^2 n \beta}{\epsilon_{\infty}}
\end{equation}
is the Debye-Huckel screening wavenumber. In the static limit $\omega \rightarrow 0$ we have that $y_{1}=y_{2}=\sqrt{\beta/2m}\hbar Q/2$, Eq. (\ref{DFreala})
yields

\begin{equation}\label{DFrealb}
\epsilon_{1}(Q, 0, T)=\epsilon_{\infty}\left[1+2\left(\frac{2 m}{\beta}\right)^{1/2}\frac{k_{DH}^2}{\hbar Q^3}D(y_{1})\right]~.
\end{equation}

In the long-wavelength limit $Q \rightarrow 0$, we have that $D(y_{1})\approx y_{1}=\sqrt{\beta/2m}\hbar Q/2$. From Eq. (\ref{DFrealb}) we obtain

\begin{equation}\label{DFrealc}
\epsilon_{1}(Q, 0, T)=\epsilon_{\infty}\left[1+\frac{k_{DH}^2}{ Q^2}\right]~.
\end{equation}

In this high-temperature limit, the exponential factors in Eq. (\ref{Imaggeralbb}) are small, so we can expand the argument of the logarithm and we obtain

\begin{equation}\label{Imaggeralb}
\epsilon_{2}(Q,\omega, T)=\epsilon_{\infty}\frac{k_{DH}^2}{\hbar Q^3}
\left[\exp(-y_{2}^2)-\exp(-y_{1}^2)\right] ~.
\end{equation}

In the static limit $\omega \rightarrow 0$, $\epsilon_{2}(Q, 0)=0$.

In the low-temperature limit, $T \rightarrow 0$ or $\beta \rightarrow \infty$, we can use the asymptotic representation of the Dawson's integral

\begin{equation}
D(y) \approx \frac{1}{2 y}\left(1+\frac{1}{ 2 y^2}\right)~.
\end{equation}

 After some simple calculation we obtain from Eq. (\ref{DFreal})

\begin{equation}\label{DFreald}
\epsilon_{1}(Q, \omega, T)=\epsilon_{\infty}\left[1-\frac{\omega_{pl}^2}{\omega ^2} \left(1-\frac{3 Q^2}{m^{*}\beta \omega^2}\right)\right]~.
\end{equation}
where

\begin{equation}\label{Debye}
\omega_{pl}^2=\frac{4 \pi e^2 n }{\epsilon_{\infty}}
\end{equation}
is the plasma frequency. In the long-wavelength limit $Q \rightarrow 0$, we obtain

\begin{equation}\label{DFreale}
\epsilon_{1}(0, \omega, T)=\epsilon_{\infty}\left[1-\frac{\omega_{pl}^2}{\omega ^2} \right]~.
\end{equation}

For the imaginary part we obtain for $T \rightarrow 0$

\begin{equation}\label{Imaggeralb}
\epsilon_{2}(Q,\omega, )=\frac{2 e^2 (m^*)^2}{\hbar^3 Q^3}\omega  ~.
\end{equation}

\section{Numerical results}

\begin{figure}
\begin{center}
\includegraphics[scale=.40]{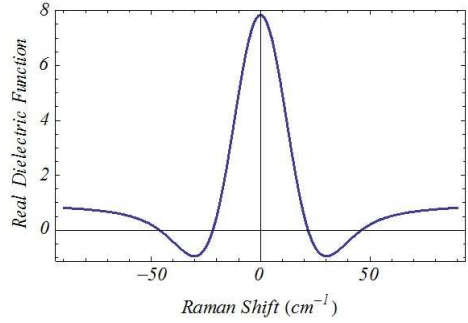}

\caption{The real dielectric function for a GaAs sample with n$=10^{16}$ $cm^{-3}$,  $Q=2.5$  $10^5 cm^{-1}$,  for $T=50$ K.}
\label{figura1}
\end{center}
\end{figure}

\begin{figure}
\begin{center}
\includegraphics[scale=.40]{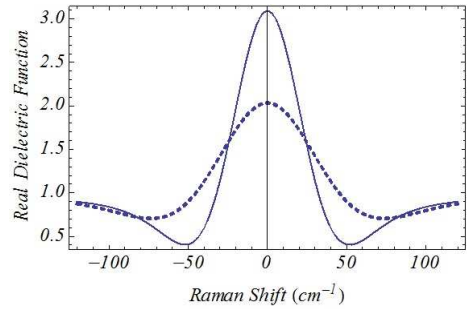}

\caption{The real dielectric function for a GaAs sample with n$=10^{16}$ $cm^{-3}$, $Q=2.5$  $10^5 cm^{-1}$,  for $T=150$ K (solid curve) and $T = 300$ K (dashed curve).}
\label{figura2}
\end{center}
\end{figure}

In order to complete our work we need obtain an analytical expression for chemical potential. At this point, we can use the called Pade approximation to obtain an analytic approximation for $\beta \mu$ which is given by \cite{Haug}

\begin{equation}\label{Eq32}
\beta \mu=\ln(\nu)+K_{1}\ln(K_{2}\nu+1)+ K_{3}\nu~,
\end{equation}
where $\nu=n/n_{o}$ with

\begin{equation}
n_{o}=\frac{1}{4}\left(\frac{2m}{\pi \hbar^2 \beta}\right)^{3/2}~,
\end{equation}
and $K_{1}=4.896685$, $K_{2}=0.04496457$ and $K_{3}=0.133376$, The Eq.(\ref{Eq32}) is a good approximation for the range $-\infty < \beta \mu \leq 30$.

\begin{figure}
\begin{center}
\includegraphics[scale=.40]{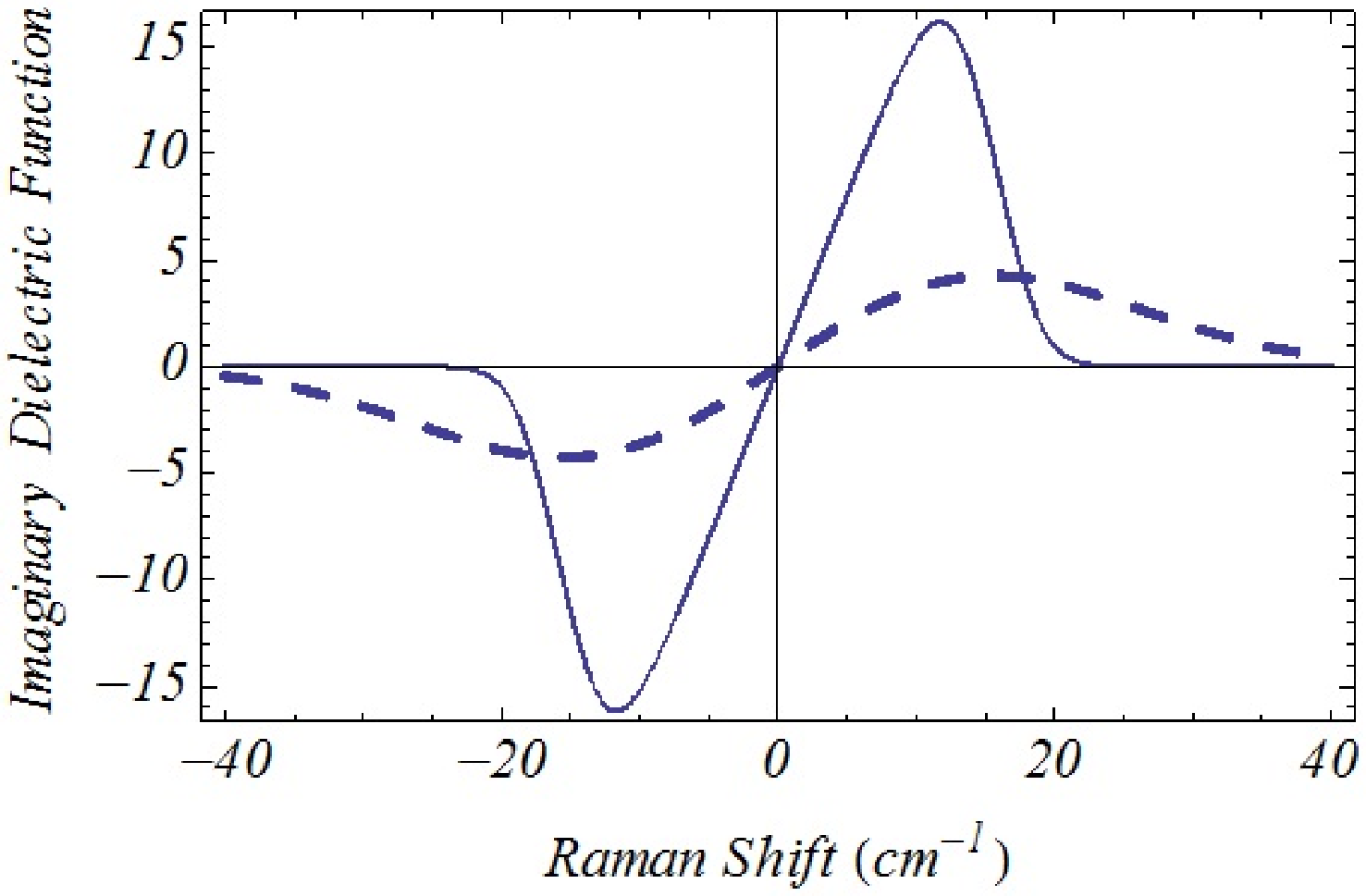}

\caption{The imaginary dielectric function for a GaAs sample with n$=10^{16}$ $cm^{-3}$, $Q=2.5$  $10^5 cm^{-1}$,  for $T=5$ K (solid curve) and $T = 50$ K (dashed curve).}
\label{figura3}
\end{center}
\end{figure}

\begin{figure}
\begin{center}
\includegraphics[scale=.40]{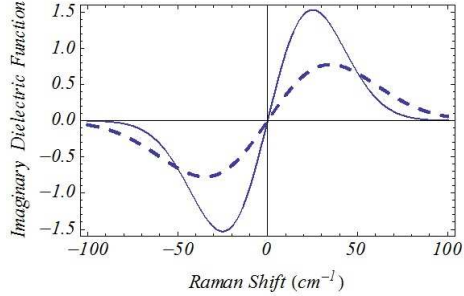}

\caption{The imaginary dielectric function for a GaAs sample with n$=10^{16}$ $cm^{-3}$, $Q=2.5$  $10^5 cm^{-1}$,  for $T=150$ K (solid curve) and $T = 300$ K (dashed curve).}
\label{figura4}
\end{center}
\end{figure}

The real part of the dielectric function,calculated from  Eq.(\ref{DFreal})  as a function of frequency, $\epsilon_{1}(\omega)/\epsilon_{\infty}$ are shown in Figure 1 (for $T= 50 K$) and Figure 2 (for $T= 150 K$ and $T= 300 K$), for doped n-type GaAs with $n=10^{16}$ $cm^{-3}$, $Q= 10^5$ $cm^{-1}$. Other parameters are $\epsilon_{\infty}=10.5$, $m^{*}=0.067 m_{o}$.

The frequency dependence of the imaginary part, $\epsilon_{2}(\omega)/\epsilon_{\infty}$, calculated after Eq.(\ref{Imaggeral}), are displayed in Figure 3  (for $T= 5 K$ and $T= 50 K$) and Figure 4 (for $T= 150 K$ and $T= 300 K$) for the same values.

We can see that the real part is an even function in frequency, $\epsilon_{1}(-\omega)=\epsilon_{1}(\omega)$, and the imaginary part is an odd function
$\epsilon_{2}(-\omega)=-\epsilon_{1}(\omega)$, a well-known property of dielectric function. Furthermore, $\epsilon_{1}(\omega)$ always approachs $1$ at large $\omega$. We can see too that for small $T$ values $\epsilon_{2}(\omega)$ has triangular form.

\section{Raman scattering}

To show the feasibility and advantages of the equations calculated in this work, now is presented an application for light scattering from single-particle electrons in semiconductors.

According to the standard theory of Raman scattering the cross section is directly proportional to the Fourier transform in space and time of the electron density-density correlation function \cite{Cardona}. But, according to the fluctuation-dissipation theorem, it results related to the frequency $\omega$ and wavevector $Q$ dielectric function $\epsilon (Q,\omega)$ in this form \cite{Neto2}

\begin{equation}\label{Raman}
\frac{d^{2}\sigma}{d\Omega d\omega} \sim [1-\exp{(-\beta\hbar\omega)}]^{-1} Im\left[\frac{-1}{\epsilon(\vec{Q},\omega)}\right]
\end{equation}
where $Im$ stands for imaginary part, $d \omega$ is the frequency interval in the spectrum and $d \Omega$ is the element of solid angle subtended by the optical window in the experimental apparatus.

\begin{figure}
\begin{center}
\includegraphics[scale=.40]{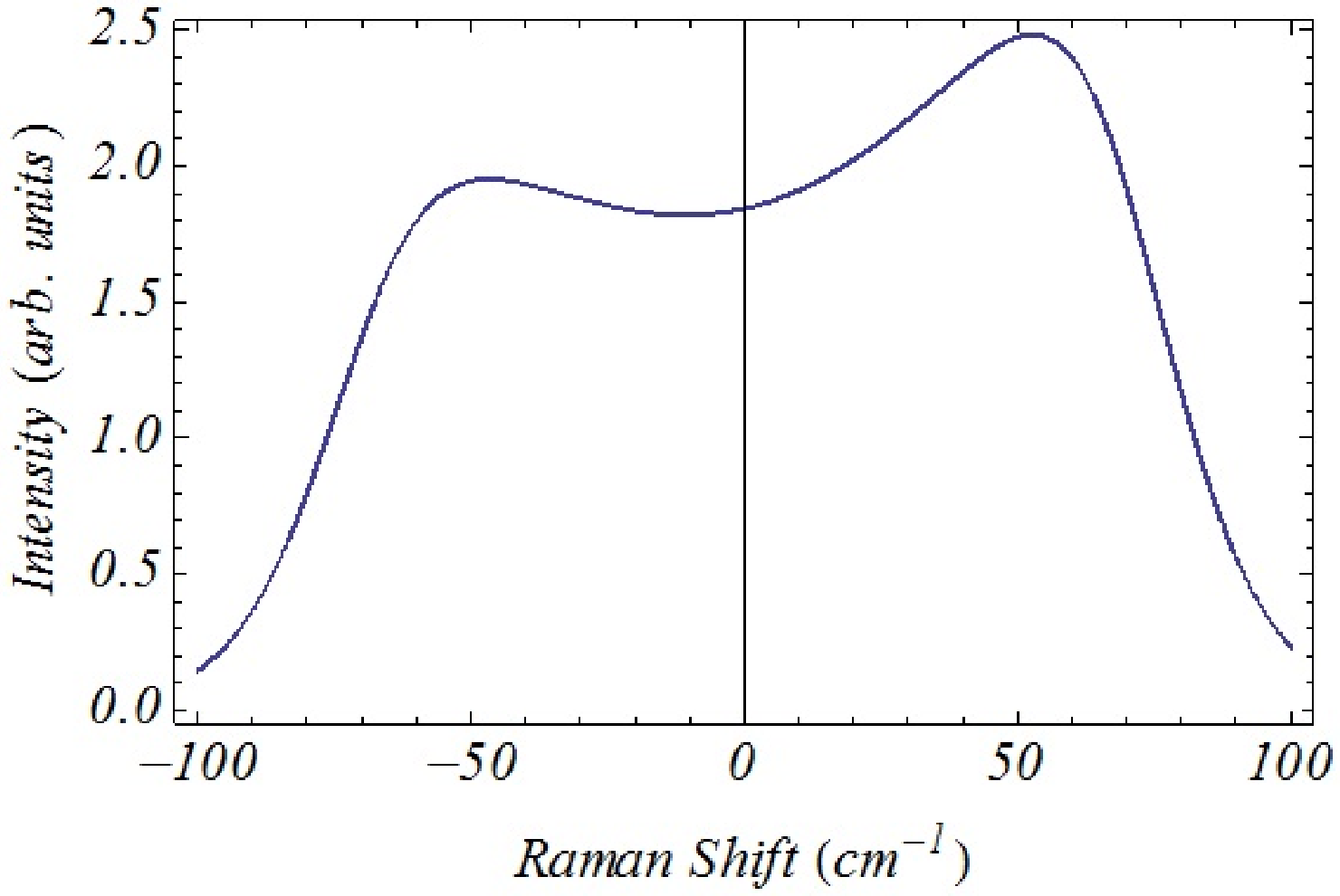}

\caption{Raman spectrum GaAs for a GaAs sample with n$=10^{16}$ $cm^{-3}$, $Q=2.5$  $10^5 cm^{-1}$,  for $T=300$ K .}
\label{figura5}
\end{center}
\end{figure}

\begin{figure}
\begin{center}
\includegraphics[scale=.40]{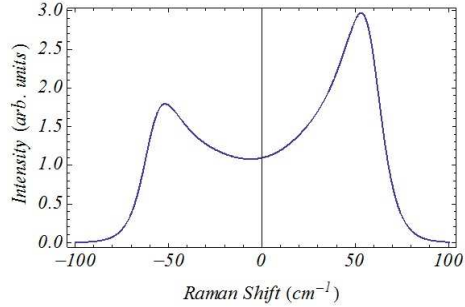}

\caption{Raman spectrum GaAs for a GaAs sample with n$=10^{16}$ $cm^{-3}$, $Q=2.5$  $10^5 cm^{-1}$,  for $T=150$ K .}
\label{figura6}
\end{center}
\end{figure}

Figures 5 and 6 shows the Raman spectra calculated from Eq.(\ref{Raman}), from doped n-type GaAs with $n=10^{16}$ $cm^{-3}$ and $Q=2.3\times 10^{5}$ $cm^{-3}$ for $T= 150 K$ (Figure 5) and $T= 300 K$ (Figure 6).

We see that with increasing temperatures there are broadenings to the plasmon peaks. At the smaller temperatures we can see well-defined plasmon peak. As the temperature increases we see clearly the damping of the plasmon line which characterize the Landau damping. This effect was experimentally demonstrated in semiconductor by Mooradian \cite{Mooradian}.

In conclusion, we have presented a derivation of explicit expression for Lindhard dielectric function for electrons in solid state plasma which is useful at finite temperature. The application performed for n-type GaAs in then range $T= 5 K$ to $T= 300 K$ highlights the usefulness of the expression obtained.

\end{document}